\documentclass[lettersize,journal]{IEEEtran}
\usepackage{amsmath,amsfonts,amssymb}
\usepackage[noend]{algorithmic}
\usepackage{algorithm}
\usepackage{array}
\usepackage[caption=false,font=normalsize,labelfont=sf,textfont=sf]{subfig}
\usepackage{textcomp}
\usepackage{stfloats}
\usepackage{url}
\usepackage{verbatim}
\usepackage{graphicx}
\usepackage{cite}
\usepackage{booktabs} 
\usepackage{multirow}
\usepackage{xcolor}
\usepackage[super]{nth}
\usepackage{tikz}
\newcommand*\circled[1]{\tikz[baseline=(char.base)]{
            \node[shape=circle,draw,inner sep=0.2pt] (char) {#1};}}
\newcommand*\circledB[1]{\tikz[baseline=(char.base)]{
            \node[shape=circle,fill,inner sep=0.2pt] (char) {\textcolor{white}{#1}};}}
\usepackage{soul}
\usepackage{enumitem}

\usepackage[switch]{lineno}

\begin{document}


\title{MDTransformer: A Hardware-Software Co-Design of Mode-Division Photonic Transformer Accelerator with Inverse-Designed Coherent Crossbar}

\author{Solomon Micheal Serunjogi$^*$, Rachmad Vidya Wicaksana Putra$^*$,~\IEEEmembership{Member,~IEEE,} Ayat Taha, Muhammad Shafique,~\IEEEmembership{Senior Member,~IEEE}, and Mahmoud Rasras,~\IEEEmembership{Senior Member,~IEEE} 
\thanks{Solomon Micheal Serunjogi and Ayat Taha are with Photonic Research Lab (PRL), Division of Engineering, New York University (NYU) Abu Dhabi, United Arab Emirates; 
(e-mail: sms10215@nyu.edu, aat9458@nyu.edu). \\
\indent Rachmad Vidya Wicaksana Putra is with eBRAIN Lab, Division of Engineering, New York University (NYU) Abu Dhabi, United Arab Emirates; 
(e-mail: rachmad.putra@nyu.edu). \\
\indent Muhammad Shafique is the Director of eBRAIN Lab, Division of Engineering, New York University (NYU) Abu Dhabi, United Arab Emirates; 
(e-mail: muhammad.shafique@nyu.edu). \\
\indent Mahmoud Rasras is the Director of Photonic Research Lab (PRL), Division of Engineering, New York University (NYU) Abu Dhabi, United Arab Emirates (UAE); 
(e-mail: mrasras@nyu.edu). \\
$^*$ Equal contributions.}
}


\maketitle


\begin{abstract}
Recently, photonic transformer accelerators (PTAs) have successfully achieved significant speedup and energy efficiency improvements over electronic accelerators for expediting Transformer inference. 
However, state-of-the-art rely on expensive multi-wavelength light generation and large dot-product units due to active phase-shifter components, thus making their approach inefficient and impractical. 
To address this, we propose \textit{\textbf{MDTransformer}}, a novel hardware-software co-design of PTA based on mode-division optical dataflow and operations.
Specifically, \textit{MDTransformer} performs complex matrix operations using spatial-mode interference, that leverages the inverse-designed multi-mode couplers, crossings, and Mach-Zehnder IQ modulators into a compact \textit{mode-division photonic tensor core (MPTC)}, capable of executing matrix multiplications in the optical domain. 
Its each guided mode (i.e., TE$_0$-TE$_3$) acts as an independent computational lane, enabling four-fold parallelism-per-waveguide without spectral filtering or free-spectral-range limitations. 
Moreover, its coherent detection and IQ modulation jointly encode amplitude and phase, realizing complex-valued arithmetic for full-range operations in transformers. 
\textit{MDTransformer} offers analog multiplication with sub-4-bit effective precision and inter-modal crosstalk below -30 dB. 
Its inverse-designed approach also offers scalable and full compatibility with single-laser continuous-wave operation at 1550 nm. 
Experimental results show that \textit{MDTransformer} achieves 40.4\% area reduction, 63.6\% power saving, 40.6\% energy saving, and comparable latency over the state-of-the-art PTA across different workloads (i.e., DeiT-Tiny/Small/Base and BERT-Base/Large). 
These results show that \textit{MDTransformer} offers a practical solution for high-performance and energy-efficient transformer-based systems. 
\end{abstract}


\section{Introduction}
\label{Sec_Intro}

\begin{figure}[t]
\centering
\includegraphics[width=\linewidth]{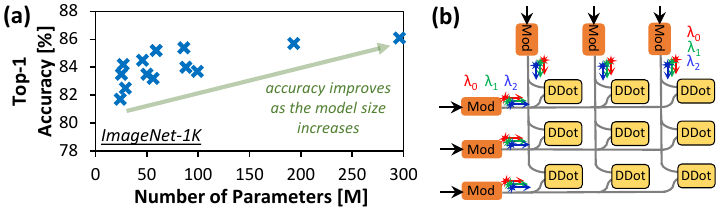}
\vspace{-0.6cm}
\caption{\textbf{(a)} Transformer networks typically improve their performance at the cost of larger memory footprint; based on data from~\cite{Ref_Han_SurveyViT_TPAMI22}.
\textbf{(b)} The state-of-the-art photonic tensor core for Transformer acceleration based on dynamically-operated dot-product (DDot) unit from the LT accelerator~\cite{Ref_Zhu_LightTrans_HPCA24}.}
\label{Fig_Trends}
\vspace{-0.3cm}
\end{figure}

Transformer-based networks~\cite{Ref_Vaswani_Attention_NIPS17}, such as Large Language Models (LLMs) and Vision Transformers (ViTs), have demonstrated state-of-the-art performance (e.g., accuracy) for solving diverse machine learning (ML) tasks, e.g., natural language processing (NLP) and computer vision~\cite{Ref_Dosovitskiy_Transformers_ICLR21, Ref_Touvron_Transformers_ICML21, Ref_Khan_SurveyViT_CSUR22}, thereby paving the way toward artificial general intelligence (AGI)~\cite{Ref_Yenduri_AGIsurvey_Access25}. 
This state-of-the-art performance comes at higher computational and memory costs as shown in Fig.~\ref{Fig_Trends}(a), 
thereby leading to huge power/energy consumption~\cite{Ref_Han_SurveyViT_TPAMI22}.
This condition limits the wide adoption of Transformer models in diverse application use-cases.
Toward this, specialized electronic accelerators for Transformer inference have been developed~\cite{Ref_Wang_Spatten_HPCA21, Ref_Zhou_Transpim_HPCA22, Ref_You_Vitcod_HPCA23}.
However, such conventional accelerators face challenges as transistor circuits hit the limits of Dennard scaling~\cite{Ref_Sunny_SurveyPhot4DL_JETC21}, leading to their diminishing return of performance efficiency (e.g., slower performance gains and increased power dissipation-per-unit area).

Recent works have proposed optical-based integrated circuits to expedite neural network (NN) inference, exploiting ultra-high speed and low energy nature of optical-based computation, known as \textit{photonic accelerators}~\cite{Ref_Shiflett_Albireo_ISCA21, Ref_Shastri_Photonics4AI_NaturePhot21, Ref_Yin_Simphony_DAC25}. 
They typically leverage optical components such as Micro-Ring Resonator (MRR)~\cite{Ref_Tait_PhotWeightBanks_SciRep17}\cite{Ref_Sunny_CrossLight_DAC21}, Mach-Zehnder Interferometer (MZI)~\cite{Ref_Shen_DLwCoherentPhot_NaturePhot17}, and Phase Change Material (PCM)~\cite{Ref_Feldmann_PCMcrossbar_Nature21} for designing a \textit{photonic tensor core (PTC)}. 
However, these works mainly target convolutional neural networks (CNN) acceleration~\cite{Ref_Zhu_LightTrans_HPCA24}, exposing the need for studies that target transformer acceleration. 
Therefore, \textit{\textbf{the targeted problem} in this work is how can we develop a high performance and energy-efficient photonic transformer accelerator (PTA)?} 
A solution to this problem may enable a practical PTA design for diverse application use-cases.

\subsection{State-of-the-Art PTAs and Their Limitations}
\label{Sec_Intro_SOTA}

Most of PTA designs employ PTC based on MZI, MRR banks~\cite{Ref_Li_SPRINT_TPDS22, Ref_Li_SPACX_HPCA22, Ref_Afifi_TRON_GLSVLSI23, Ref_Afifi_LLMoptical_GLSVLSI25, Ref_Afifi_ASTRA_TECS25}, and PCM crossbars~\cite{Ref_Li_MERIT_TSUSC25}. 
They statically store operands on the optical components for computation, hence they suffer from slow operand mapping and programming. 
Recently, the Lightening-Transformer (LT) accelerator~\cite{Ref_Zhu_LightTrans_HPCA24} has been proposed. 
It inspires further studies in reconfigurability aspect~\cite{Ref_Zhu_ENlighten_arXiv25} and digital-to-analog converter (DAC) optimization~\cite{Ref_Li_HyAtten_DATE25}\cite{Ref_Chang_PDAC_DAC25}. 
LT improves performance efficiency of Transformer inference over other PTA designs by employing PTC with dynamic operations of full-range input operands through \textit{dynamically-operated dot-product (DDot)} unit; see Fig.~\ref{Fig_Trends}(b). 
Hence, it eliminates slow operand mapping/programming and making it the state-of-the-art PTA design. 
Despite their benefits, \textit{all these works still have the following critical limitations}.
\begin{itemize}[leftmargin=*]
    \item They typically employ multiple wavelengths for operations based on wavelength-division multiplexing (WDM), to achieve highly parallel multiply-accumulate (MAC) operations~\cite{hamerly2024netcast}\cite{li2025hybrid}. 
    However, their reliance on finely-spaced resonant filters and dispersion-limited channels imposes scalability bottlenecks, as generating multiple wavelengths consumes huge area and power/energy and is often done in a strongly nonlinear medium such as SiN. 
    \item The free-spectral-range (FSR) of MRRs restricts the number of usable wavelengths, while temperature-dependent resonance drift and fabrication non-uniformity demand active thermal control and calibration overhead~\cite{biasi2024photonic,tait2016microring}. 
    \item Coherent operation across many wavelength lanes requires precise optical phase alignment and stabilization, thereby adding power and system complexity for a practical solution~\cite{banerjee2022characterizing}\cite{totovic2022programmable}.
    \item State-of-the-art accelerators relies on \textit{Micro\_comb (MC)}-based wavelength generator, \textit{Mach-Zehnder Modulator (MZM)}, and \textit{phase-shifter (PS)}-based PTC, which are area and power hungry. 
    Therefore, they impose scalability and efficiency challenges when designing area- and power-efficient PTA architecture.
\end{itemize} 
To show the limitations of state-of-the-art and related research challenges, we conduct a case study in Section~\ref{Sec_Intro_StudyChallenges}.

\subsection{Case Study and Related Research Challenges}
\label{Sec_Intro_StudyChallenges}

\begin{figure}[t]
\centering
\includegraphics[width=\linewidth]{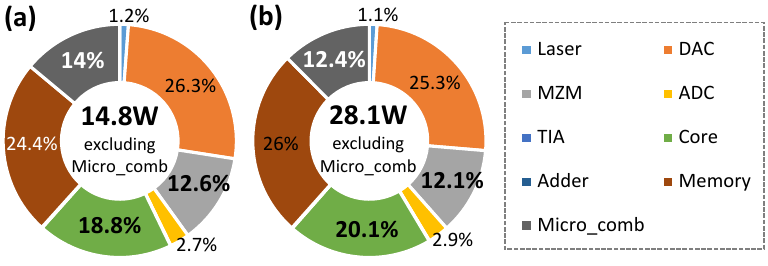}
\vspace{-0.6cm}
\caption{Area breakdown of the state-of-the-art 4-bit LT accelerators: \textbf{(a)} LT-Base and \textbf{(b)} LT-Large, showing the contributions of different modules.}
\label{Fig_CaseStudy}
\vspace{-0.4cm}
\end{figure}

We study the impact of different modules in the state-of-the-art 4-bit LT accelerators (i.e., LT-Base and LT-Large)~\cite{Ref_Zhu_LightTrans_HPCA24} on area and power consumption using its open-source codes from the original authors. 
The experimental results are shown in Fig.~\ref{Fig_CaseStudy}, from which we draw the following key observations.
\begin{itemize}[leftmargin=*]
    \item Micro\_comb, MZM, and PS-based PTC jointly occupy 45.5\% area in LT-Base and 44.6\% area in LT-Large, highlighting their dominant area consumption.
    \item Both LT accelerators incur high power consumption, even without Micro\_com. 
    It is particularly inefficient for meeting diverse possible power-constrained computing systems. 
    For instance, embedded AI systems typically require about 5W max. power envelope, which is difficult to meet with the existing solutions.
    \item Employing a Micro\_comb module, which resides in a separate physical chip, comes with additional complexity and non-trivial challenges since it requires a highly precise control in different aspects, including optical stabilization and power distribution. 
\end{itemize}

These observations expose the following \textit{key research challenges} to address for providing a practical PTA design solution.
\begin{itemize}[leftmargin=*]
    \item The use of Micro\_comb should be avoided to minimize design complexity and inter-chip communication challenges. 
    Hence, its laser generation functionality should be replaced with an efficient alternative solution.
    \item The area of laser generator, modulator, and PTC should be optimized to significantly reduce area, and hence minimizing power and energy consumption.
    \item PTA architecture and dataflow should be synergistically designed to  exploit to maximize the performance and efficiency benefits offered by the optical-based processing. 
\end{itemize}

\subsection{Our Novel Contributions}
\label{Sec_Intro_Novelty}

To address the targeted problem and related challenges, we propose \textbf{MDTransformer}, \textit{a novel hardware-software (HW-SW) co-design of photonic transformer accelerator (PTA) that leverages the spatial \textbf{Mode-Division Multiplexing (MDM)}, \textbf{inverse-designed coherent crossbar}, and \textbf{IQ modulation} to enable a practical solution for high-performance and energy-efficient transformer-based systems}. 
It is also the first work that leverages MDM and inverse design concepts for designing PTA. 
It employs the following key ideas. 
\begin{itemize}[leftmargin=*]
    \item \textbf{Leveraging Spatial MDM for Photonic Computing (Section~\ref{Sec_MDT_MDM}).}
    It employs MDM to enable on-chip photonic parallelism by employing efficient Mode-based Multiplexer or Demultiplexer (MUX/DEMUX), which distributes, encodes, and routes a single optical source across multiple guided spatial channels. 
    \item \textbf{MDOT: Mode-Division Dot-Product Unit (Section~\ref{Sec_MDT_MDOT}).} 
    It aims to perform dot-product operation that represents multiplication of two full-range operands in multiple modes at sigle frequencies, thereby enabling dynamically-operated processing element (PE) for higher-level architecture hierarchy (i.e., PTC).
    \item \textbf{MPTC: Mode-Division Photonic Tensor Core (Section~\ref{Sec_MDT_MPTC}).}
    It aims to efficiently accelerate general matrix multiplication (GEMM) by leveraging MDOT, Mode-based MUX/DEMUX, IQ modulator, and crossings in a crossbar array fashion. 
    \item \textbf{Architecture System Design (Section~\ref{Sec_MDT_MPTC}).}
    It aims to develop the architecture system of \textit{MDTransformer} accelerator, by integrating multiple MPTCs and the supporting digital circuits (e.g., on-chip memory). Furthermore, a dataflow pattern is also developed to maximize the benefits of the \textit{MDTransformer} architecture.  
\end{itemize}

\textbf{Key Results:} 
We evaluate \textit{MDTransformer} through functional simulation using Tidy3D~\cite{flexcompute_tidy3d}, as well as hardware evaluation (e.g., area, power, energy, and latency) using the state-of-the-art PTA hardware simulator from~\cite{Ref_Zhu_LightTrans_HPCA24}.
Furthermore, our design is also under fabrication.
Experimental results show that, \textit{MDTransformer} offers 4-bit effective precision for multiplication, low inter-modal crosstalk (i.e., below -30 dB), and full compatibility with single-laser continuous-wave operation at 1550nm. 
\textit{MDTransformer} also achieves 40.4\% area reduction, 63.6\% power saving, 40.6\% energy saving, and comparable latency over the state-of-the-art PTA across different workloads (i.e., DeiT-T/S/B\footnote{DeiT-T/S/B denotes DeiT-Tiny, DeiT-Small, and DeiT-Base, respectively.} and BERT-B/L\footnote{BERT-B/L denotes BERT-Base and BERT-Large, respectively.}).

\section{Preliminaries}
\label{Sec_Prelim}

\subsection{Transformer-based Network Models}
\label{Sec_Prelim_Transformer}

A transformer-based network is formed by multiple identical blocks: \textit{encoder} and \textit{decoder} blocks.
Each block is formed by a multi-head self-attention (MA) module, a feed-forward network (FN) module, a layer normalization (LN) module, and shortcut connections~\cite{Ref_Zhu_LightTrans_HPCA24}. 
The decoder block also has cross-attention and masked self-attention modules. 
The basic encoder block can be stated as Eq.~\ref{Eq_Encoder}-\ref{Eq_Encoder2}, where $\textbf{X}_l$ is the input sequences of $l$-th layer.
Multi-head self-attention (MA) module supports $H$ self-attention heads, and each head makes the input vector into separate vectors, i.e., query (\textbf{Q}), key (\textbf{K}), and value (\textbf{V}) vectors.
The attention function between these input vectors can be state as Eq.~\ref{Eq_Attention}, where $d_k$ is the dimension of \textbf{Q} and \textbf{K}. 
\begin{equation}
  \begin{split}
    \textbf{X}_{l+1} = \textit{FN}(\textit{LN}(\hat{\textbf{X}}_{l+1}))+\hat{\textbf{X}}_{l+1} 
  \end{split}
  \label{Eq_Encoder}
\end{equation}
\begin{equation}
  \begin{split}
    \hat{\textbf{X}}_{l+1} = \textit{MA}(\textit{LN}(\textbf{X}_l))+\textbf{X}_l 
  \end{split}
  \label{Eq_Encoder2}
\end{equation}
\begin{equation}
    Atten(\textbf{Q}, \textbf{K}, \textbf{V}) = softmax\left(\frac{\textbf{Q}\textbf{K}^{\intercal}}{\sqrt{d_k}}\right) \textbf{V}
  \label{Eq_Attention}
\end{equation}

\subsection{Inverse Design in Photonic Circuits}
\label{Sec_Prelim_InverseDesign}
 
\begin{figure*}[t]
    \centering
    \includegraphics[width=\linewidth]{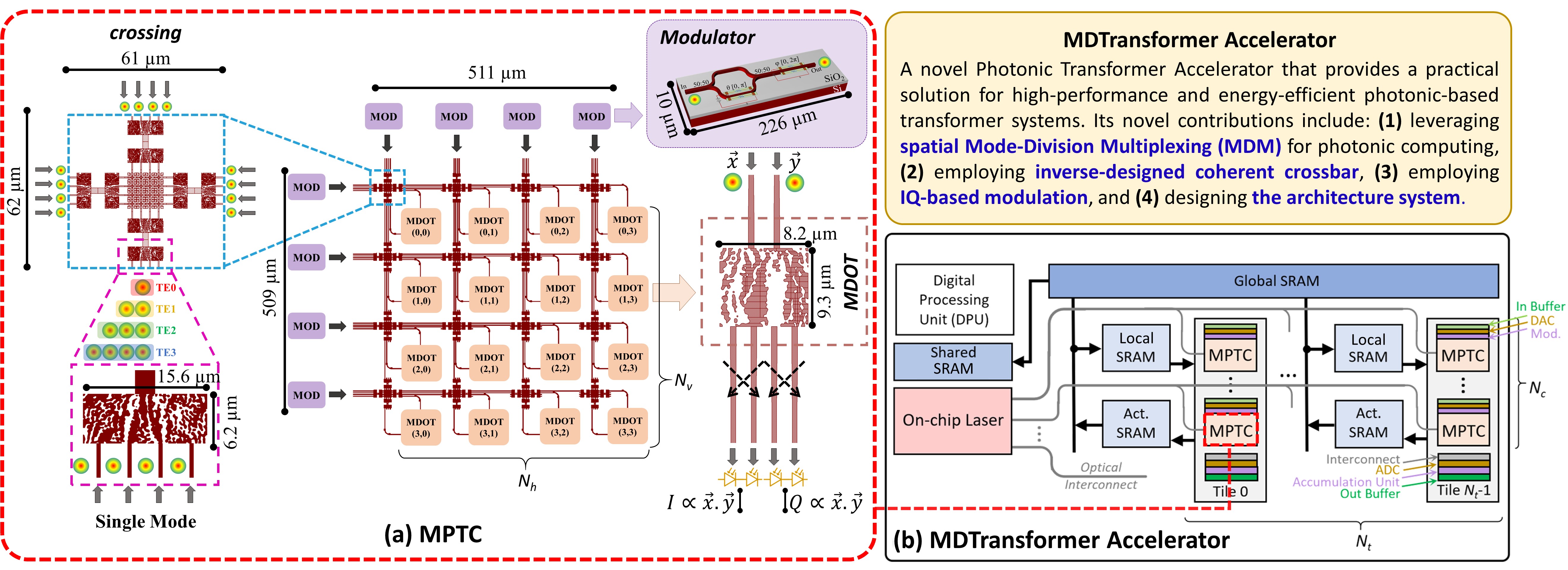}
    \vspace{-0.7cm}
    \caption{Our proposed \textit{MDTransformer} Accelerator: 
    \textbf{(a)} MPTC with crossings, modulator, and MDOT; \textbf{(b)} architecture design.}
    \label{Fig_MDT}
    \vspace{-0.3cm}
\end{figure*}

Inverse-designed components can implement complex transformations, such as filters and couplers, within an order of magnitude smaller than classical designs~\cite{ sapra2020inverse, jensen2020inverse, vercruysse2020compact}. 
Such devices maintain low loss and high modal fidelity, making them ideally suited for large-scale photonic accelerators here thousands of operations must be packed into a small area. 
As photonics moves toward ultra-dense, domain-specific optical computing, inverse design has become a promising approach for building high-performance primitives that enable massive parallelism and energy-efficient linear algebra.


\section{Our Proposed \textit{MDTransformer} Accelerator}
\label{Sec_MDT}

We develop our PTA design, called \textit{MDTransformer}, based on MDM, inverse-designed coherent crossbar, and IQ modulation (overview in Fig.~\ref{Fig_MDT}).
Details of its design is discussed in Section~\ref{Sec_MDT_MDM} - Section~\ref{Sec_MDT_Arch}.

\subsection{Leveraging Spatial Mode-Division Multiplexing for Photonic Computing}
\label{Sec_MDT_MDM}

We leverage spatial MDM to obtain an orthogonal degree of freedom for on-chip photonic parallelism~\cite{wang2022mode,liu2019arbitrarily}, thereby enabling light generation using limited number of on-chip lasers and reducing on the number of WDM channels.  
Instead of distributing computation across distinct wavelength like in the state-of-the-art works~\cite{Ref_Zhu_LightTrans_HPCA24, 
Ref_Li_HyAtten_DATE25, Ref_Chang_PDAC_DAC25}, \textit{MDM leverages multiple guided modes (e.g., TE$_0$–TE$_3$) within a single multi-mode waveguide, enabling independent and simultaneous information channels in the spatial domain}. 
Here, operand pairs are mapped onto orthogonal modal channels (e.g., TE$_0$–TE$_3$) of a multi-mode bus waveguide and processed in modular MDOT units. 

\subsection{MDOT: Mode-Division Dot-Product Unit}
\label{Sec_MDT_MDOT}

\begin{figure}[t]
    \centering
    \includegraphics[width=\linewidth]{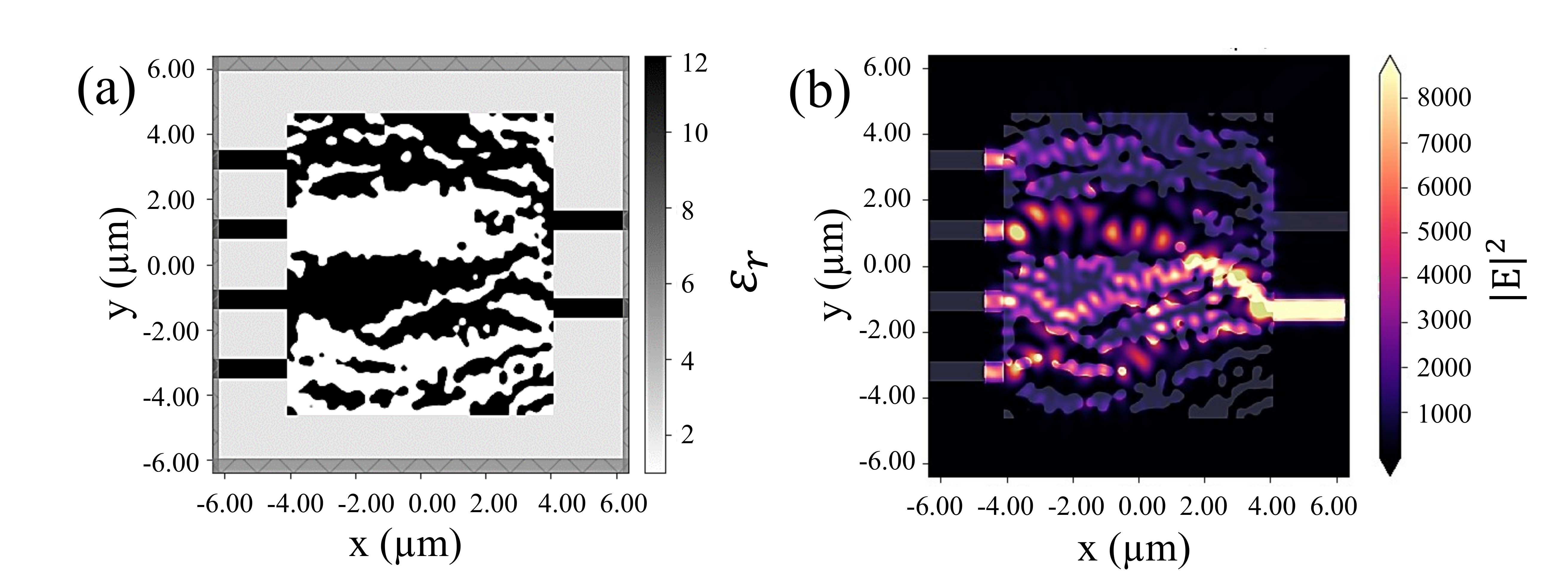}
    \vspace{-0.7cm}
    \caption{Inverse-designed MDOT design: 
    \textbf{(a)} silicon projection of the design region.
    \textbf{(b)} Intensity distribution of light traveling through the structure from the left input.}
    \label{Fig_MDOTsim}
    \vspace{-0.2cm}
\end{figure}

We propose a \textit{Mode-Division Dot-Product Unit (MDOT)} to perform a signed multiplication between two full-range mode-encoded operands through optical interference; see Fig.~\ref{Fig_MDOTsim}.
Fig.~\ref{Fig_MDOTsim}(a) shows the novel inverse-designed structure after optimization, which ensures that the coherent coupler occupies a small footprint and does not need an area- and power-hungry 90-degree phase-shifter. 
Meanwhile, Fig.~\ref{Fig_MDOTsim}(b) shows the intensity across the MDOT structure.
Here, each spatial mode is routed into a compact $8\times 8~\mu\mathrm{m}^2$ inverse-designed coherent mixer that produces four output interference states with fixed phase relationships.
Each of the four mixer outputs can be described using a linear transformation of the two incoming operands $s_A$ and $s_B$; see Eq.~\ref{Eq_MDOT}. 
It explicitly shows the $\{0^\circ,\,90^\circ,\,
180^\circ,\,-90^\circ\}$ phase basis generated at the outputs.
\begin{equation}
\label{Eq_MDOT}
\begin{split}
[E_1 & \; E_2 \; E_3 \; E_4]^{\intercal}
= \\
& \frac{1}{2}
[(s_A + j s_B) \; 
(s_A - j s_B) \;
(s_A - s_B) \;
(s_A + s_B)]^{\intercal}
\end{split}
\end{equation}

\begin{figure}[t]
    \centering
    \includegraphics[width=\linewidth]{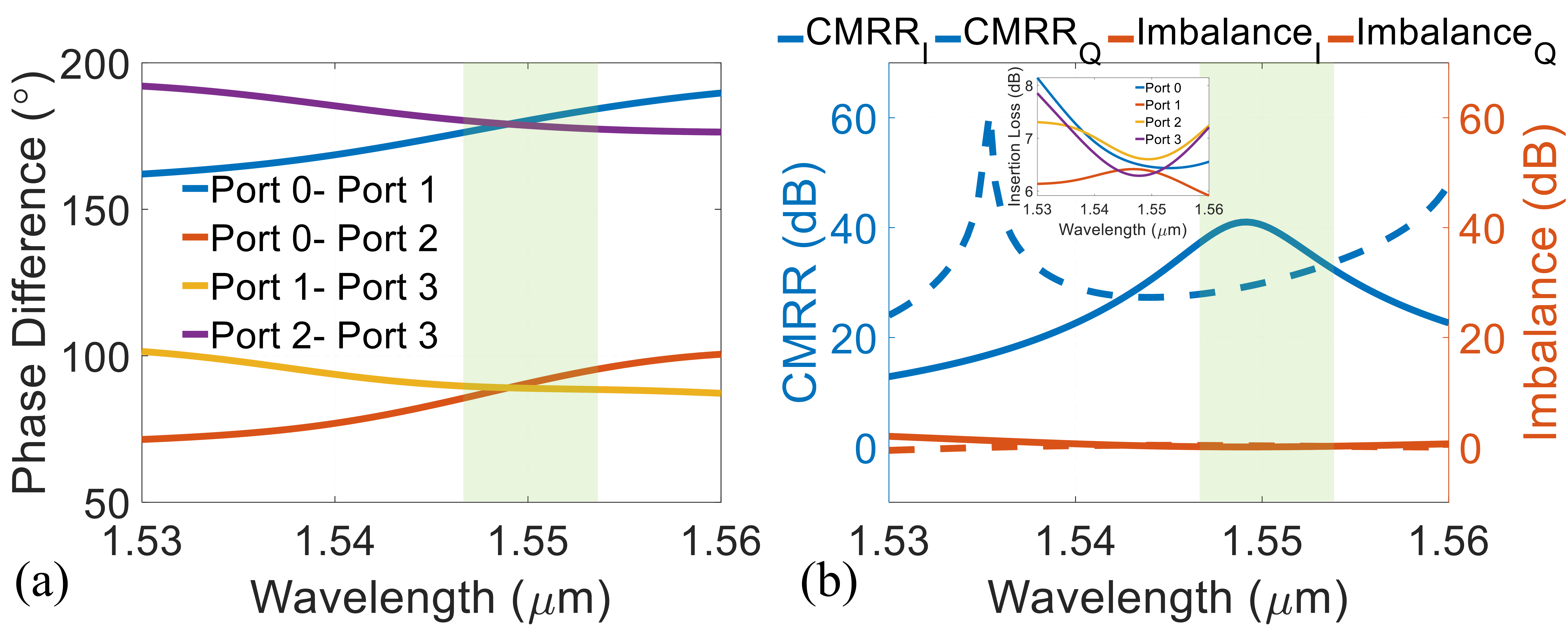}
    \vspace{-0.6cm}
    \caption{MDOT properties: 
    \textbf{(a)} simulated relative phase differences between the four output ports; and \textbf{(b)} CMRR.}
    \label{Fig_CMRR}
    \vspace{-0.2cm}
\end{figure}

Figure ~\ref{Fig_CMRR}(a) shows the simulated relative phase differences between all four output ports, taken pairwise, when a single mode is launched from the input (left side).
Adjacent port pairs (port~0–port~1 and port~2–port~3) maintain approximately $180^\circ$ phase separation across the 1530–1560~nm band, while alternating port pairs maintain roughly $90^\circ$ separation. The figure also shows a near zero port imbalance across the wavelength of interest as shown by the lightly shaded green region.
Meanwhile, the corresponding \textit{common-mode rejection ratio} (CMRR) is shown in Fig.~\ref{Fig_CMRR}(b), exceeding $30$~dB at the operating wavelength of $\lambda=1550$~nm.

\smallskip
\textbf{Bipolar Encoding and Signed Multiplication:}  
The MDOT unit accepts two bipolar NRZ symbol streams
$s_x^{(m)}(k)$ and $s_y^{(m)}(k)\in\{-1,+1\}$ derived from input bits $a_k,b_k$ via $s=1-2a$. The fields applied to the coherent mixer are:
\begin{equation}
E_x^{(m)}(k)=s_x^{(m)}e^{j\phi_x^{(m)}},\qquad
E_y^{(m)}(k)=s_y^{(m)}e^{j\phi_y^{(m)}},
\end{equation}
with $\phi_x=\pi p_k$ and $\phi_y=\pi q_k$ and $p_k,q_k\in\{0,1\}$ applied through a phase modulator. The balanced detection photocurrent for mode $m$ over $N$ symbol periods yields the mode-wise dot product:
\begin{equation}
\begin{split}
I_{PC}^{(m)}
& =
C
\int_{0}^{N\tau}
s_x^{(m)}(t)\,
s_y^{(m)}(t)\,
\cos\!\big(\Delta\phi_m(t)\big)\,dt \\
& ~\propto~
\mathbf{x}_m\!\cdot\!\mathbf{y}_m,
\label{eq:bipolar_dot}
\end{split}
\end{equation}
where $\mathbf{x}_m$ and $\mathbf{y}_m$ denote the symbol vectors carried by
mode $m$ and C is an arbitrary constant.
Summing over all supported spatial modes produces the full 
mode-division dot product:

\begin{equation}
\begin{split}
& I_{PC}^{(m)}
~\propto~
\mathbf{x}_m\!\cdot\!\mathbf{y}_m,
\;\;\; \\
& I_{PC}
=
\sum_{m=0}^{M-1} I_{PC}^{(m)}
~\propto~
\sum_{m=0}^{M-1}
\mathbf{x}_m\!\cdot\!\mathbf{y}_m.
\end{split}
\end{equation}
Therefore, the coherent MDOT unit directly performs signed multiplication and accumulation using both optical (high Q resonators) and electrical domain (capacitive dynamics) through time multiplexed integrators \cite{lam2024dynamic,ning2024photonic,babashah2019temporal}.

\subsection{MPTC: Mode-Division Photonic Tensor Core}
\label{Sec_MDT_MPTC}

We propose a novel photonic tensor-core architecture, referred to as the
\textit{Mode-Division Photonic Tensor Core (MPTC)}, for efficient general
matrix multiplication (GEMM) using a single optical carrier and multiple
orthogonal spatial modes. The MPTC combines four principal building blocks:
(i) a mode-based multiplexer/demultiplexer (MUX/DEMUX), (ii) the coherent
mode-division dot-product unit (MDOT), (iii) IQ-based complex modulation,
and (iv) compact routing elements such as crossings and couplers, all
assembled into a structured array, as shown in Fig.~\ref{Fig_MDT}(a).

Unlike prior photonic matrix engines that rely on wavelength-division
parallelism or cascaded interferometric meshes, the proposed MPTC uses
spatial modes as the primary computational lanes. This choice reduces
dependence on multiple laser wavelengths, resonance management, and
spectral routing overhead, while enabling multiple operands to propagate
within the same multimode waveguide.

\smallskip
\subsubsection{\textbf{Mode-based Multiplexer and Demultiplexer (MUX and DMUX)}}

The MUX/DEMUX is the front-end modal interface of the MPTC and is
responsible for converting single-mode input channels into a multimode
computational bus, and conversely extracting specific modes at later
processing stages. In contrast to communication-only mode multiplexers,
which are typically designed as standalone coupling elements, the
MUX/DEMUX here is designed as a computational routing primitive whose role
is to inject and recover operands inside a dense coherent dot-product
array.

\begin{figure*}[t]
    \centering
    \includegraphics[width=0.75\linewidth]{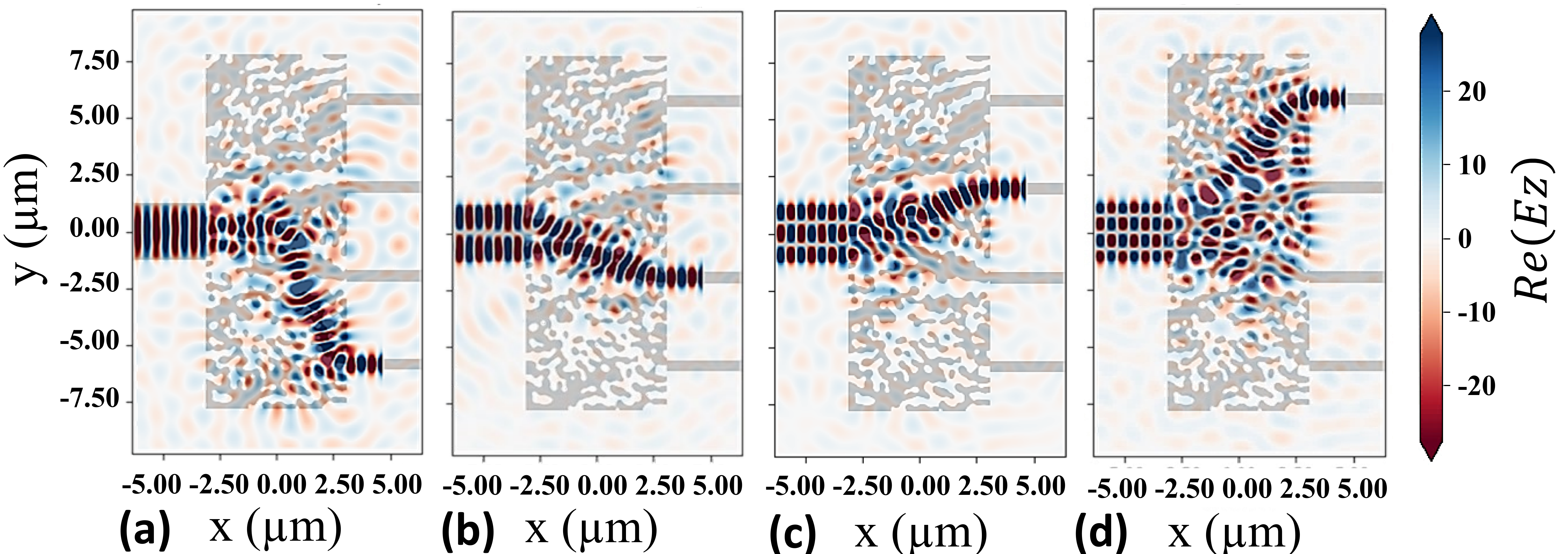}
    \vspace{-0.2cm}
    \caption{Inverse-designed mode-based MUX/DEMUX: optical field
    distributions for the four input modes, showing selective routing to
    distinct single-mode outputs. Each panel illustrates a mode-resolved
    input-to-output field transformation rather than simple power
    splitting.}
    \label{Fig_MuxDemux}
    \vspace{-0.3cm}
\end{figure*}

Fig.~\ref{Fig_MuxDemux} shows the operation of the four-mode MUX/DEMUX,
designed using full-wave inverse design in Tidy3D. The figure illustrates
the field evolution for each input mode and the corresponding selective
routing to the designated output port. Each panel should be interpreted as
a mode-resolved demonstration of selective field transformation: for each
input channel, the optical energy is redistributed so that only the target
output port carries the desired mode, while leakage to the other ports is
suppressed.

The input interface consists of four single-mode waveguides of width
$0.5~\mu$m. This width is selected to ensure robust TE$_0$ operation at
$\lambda = 1550$~nm, thereby providing a clean modal input state before
multiplexing. The spacing between adjacent input waveguides is set to
$1.5~\mu$m. This spacing is large enough to suppress unwanted evanescent
coupling between neighboring inputs, yet small enough to maintain dense
layout compatibility with the surrounding tensor-core routing network.

These four single-mode channels feed a multimode bus waveguide of width
$2.5~\mu$m. This width is chosen because it provides a practical trade-off
between modal capacity and circuit density: it is sufficiently wide to
support the first four guided TE modes (TE$_0$--TE$_3$) at 1550~nm, while
remaining narrow enough to avoid excessive crossing area, large bending
penalties, and poor array density. In other words, the selected dimensions
are not arbitrary; they arise from the joint requirement of supporting
four orthogonal modes and embedding them in a compact computational
crossbar.

To obtain the optimized freeform structure, the permittivity distribution
$\varepsilon(\mathbf{r})$ is solved through an adjoint-based gradient
descent formulation that maximizes the transmission of each target mode
into its assigned output port while penalizing leakage into all other
ports. The optimization problem is expressed in
Eq.~\ref{Eq_mdmux_opt}, where $T_{m \rightarrow p_m}$ denotes the
transmission from input mode $m$ to its designated output port $p_m$, and
$\alpha$ is a penalty factor enforcing crosstalk suppression. This
objective does not merely maximize throughput; it imposes a mode-selective
field transformation that preserves the computational meaning of each
channel.

\begin{equation}
\label{Eq_mdmux_opt}
\max_{\varepsilon(\mathbf{r})} 
\; \mathcal{F} 
= 
\sum_{m=0}^{3} 
\left(
    T_{m \rightarrow p_m}
    - 
    \alpha \!\!\sum_{\substack{n=0 \\ n\neq m}}^{3} 
    T_{m \rightarrow p_n}
\right)
\end{equation}

Physically, the optimized region acts as a compact distributed scattering
medium that directly maps one modal basis to another. This is an important
distinction from conventional asymmetric directional couplers, microring
assisted mode couplers, or MMI-based devices, where coupling is governed
by predetermined geometric interference lengths and is often less flexible
for simultaneously enforcing compactness, broadband operation, and
multi-port modal selectivity. Earlier on-chip mode-division multiplexers,
such as microring-assisted designs, demonstrated selective mode coupling
but remained tied to wavelength-sensitive routing concepts. Similarly,
ultra-compact multimode routing work has focused on bends and crossings
for dense integration. Here, by contrast, the inverse-designed MUX/DEMUX
is integrated directly into a photonic tensor-core data path, where its
purpose is not only multiplexing, but controlled operand delivery to
coherent compute nodes~\cite{liu2019arbitrarily}\cite{luo2014wdm, yang2022multi, pita2025integrated}.

\smallskip
\textbf{Fabrication-Aware Inverse Design and Constraints:}
To ensure practical manufacturability, the inverse design process
incorporates fabrication-aware constraints consistent with standard
electron-beam lithography in silicon photonics.
Specifically, a minimum feature size of $120~\mathrm{nm}$ is enforced
through spatial filtering and projection steps applied to the
permittivity distribution during optimization. This avoids the formation
of sub-resolution features and ensures that the final structure can be
faithfully fabricated without requiring additional post-processing.
Such feature-size-constrained inverse design has been widely adopted in
recent nanophotonic devices to bridge the gap between idealized
continuous permittivity optimization and binary fabrication-compatible
layouts.

\smallskip
\textbf{Modal Superposition and Decomposition Strategy:}
Although the device operates on four orthogonal modes simultaneously, the
optimization is structured using a modal decomposition approach.
Each mode transformation is treated as an independent objective, and the
total cost function is constructed as a superposition of these modal
targets, as shown in Eq.~\ref{Eq_mdmux_opt}. This ensures that each input
mode is selectively mapped to its corresponding output port without
interfering with the routing of the remaining modes.
This decomposition is physically justified by the orthogonality of the
guided modes in the multimode waveguide, allowing independent control of
each modal channel while maintaining a shared spatial structure.

\begin{figure*}[t]
    \centering
    \includegraphics[width=\linewidth]{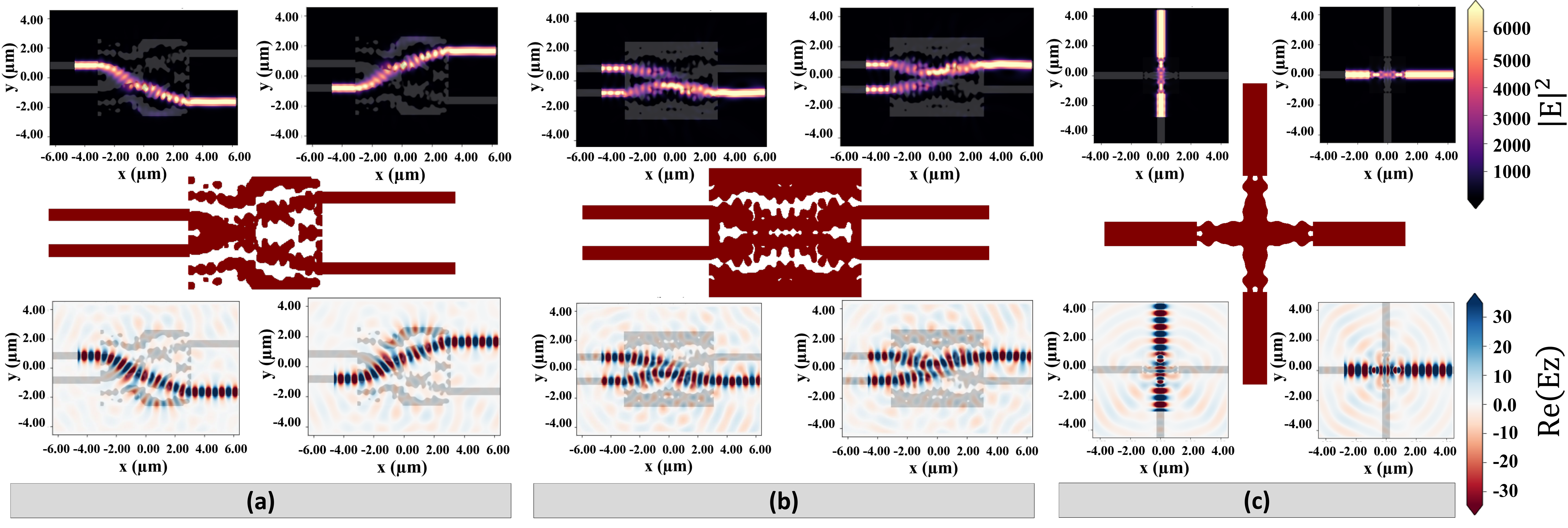}
    \vspace{-0.6cm}
    \caption{Cross-sectional optical field distributions for \textbf{(a)} scissors crossing, \textbf{(b)} 50:50 3dB coupler, and \textbf{(c)} 90$^\circ$ waveguide crossing.}
    \label{Fig_Devices}
    \vspace{-0.2cm}
\end{figure*}

\smallskip
\textbf{Reciprocity and Forward–Adjoint Consistency:}
The optimization leverages electromagnetic reciprocity, whereby the
adjoint simulation corresponds to exciting the device from the output
ports and propagating fields backward.
Consistency between forward and adjoint field distributions ensures that
the optimized structure satisfies both excitation and collection
conditions simultaneously. In practice, this guarantees that the device
performs equivalently under forward multiplexing and reverse
demultiplexing operation, which is essential for its dual role within the
MPTC architecture.

\smallskip
\textbf{Output Mode Engineering and Power Capture Efficiency:}
In order to improve power transfer from the multimode region into the output
waveguides, the single-mode output ports are intentionally widened beyond
the nominal $0.5~\mu$m width.
Specifically, the outputs are expanded to approximately $0.75~\mu$m
before being adiabatically tapered back to standard single-mode
dimensions. This local widening improves mode overlap between the
transformed field distribution and the guided mode of the output
waveguide, thereby enhancing coupling efficiency and reducing scattering
loss.
Such taper-assisted mode matching is critical in inverse-designed
structures, where the output field profile may not perfectly match the
fundamental mode of a narrow waveguide without additional impedance
matching.

\smallskip
The field distributions in Fig.~\ref{Fig_MuxDemux} also clarify why the
inverse-designed approach is needed. For each launched mode, the structure
does not simply split power; it redistributes phase and amplitude across a
freeform subwavelength region so that the desired output field emerges at
one specific port while the remaining ports are suppressed. This
mode-resolved routing behavior is exactly what is required in the MPTC:
at each downstream computational cell, one selected mode must be exposed
to the coherent multiplier, while the remaining modes must continue
propagating with minimal disturbance.

In benchmarking terms, recent inverse-designed mode-division devices have
demonstrated that compact mode multiplexers can significantly outperform
conventional mode-routing footprints, for example through five-mode
inverse-designed MDM devices with a reported footprint of
$16 \times 7~\mu$m$^2$ and measured crosstalk below approximately
$-11$~dB, as well as recent scalable mode demultiplexers with sub-1~dB
loss and crosstalk below approximately $-13$~dB at 1550~nm. Dense
multimode routing elements such as $8 \times 8~\mu$m$^2$ crossings have
also been reported for three-mode photonic circuits. Our design inherits
the compactness philosophy of these works but targets a different system
problem: rather than building a communication link, the present MUX/DEMUX
is dimensioned and optimized as the operand-injection and mode-selection
interface for a coherent photonic tensor core~\cite{liu2019arbitrarily}\cite{pita2025integrated}\cite{li2025ultra}.

Additional supporting components, including the \textit{scissors crossing},
\textit{50:50 3 dB coupler}, and \textit{90$^\circ$ waveguide crossing},
are presented in Fig.~\ref{Fig_Devices}(a)--(c), respectively. These
elements are used to construct the routing network surrounding the MPTC.
The processing pipeline begins with a continuous-wave input field that is
first split using integrated power splitters and 50:50 couplers. These
peripheral components also form the basis of other circuit blocks such as
IQ modulators and mode-scissors crossings, enabling flexible routing of
optical data throughout the processor.

After splitting, the optical field is expanded into the multimode bus
waveguide and encoded into one of the four orthogonal TE modes used by
the MD-Transformer. The inverse-designed MUX/DEMUX then maps each input
modal profile onto a unique single-mode output port at 1550~nm. This
provides clean modal separation, low inter-mode crosstalk, and a compact
footprint suitable for dense dot-product arrays.

Overall, the mode-based MUX/DEMUX forms the front-end interface of the
MD-Transformer, enabling parallel spatial-mode encoding, selective
demultiplexing, and physically structured delivery of operands into the
downstream coherent processing core.

\smallskip
\subsubsection{\textbf{IQ-Based Complex Modulator}}

To enable full complex-valued encoding of optical operands, each modal channel incorporates a compact IQ-modulator. Two complementary devices are used:  
(i) a single-input intensity I modulator for amplitude control, and  
(ii) Q modulator for phase control.  
Each input channel employs an IQ modulator analogous to that in~\cite{rahimi2024realization}, but modified to operate at high speeds of 25Gb/s. 
The amplitude branch is defined by the normalized MZI power transfer in Eq.~\ref{eq:mzi_amp}, and the quadrature branch applies an additional phase-shift as in Eq.~\ref{eq:mzi_phase}. 
Here, $A,B,D,E$ are empirical calibration coefficients from device-level measurements.
\begin{equation}
P_{\mathrm{MZI}}(I_A)
=\tfrac{1}{2}
+\tfrac{1}{2}
\cos\!\left(
A I_A^2 + B I_A + \phi_A
\right),
\label{eq:mzi_amp}
\end{equation}
\begin{equation}
\phi_{\mathrm{PS}}(I_Q)
= D I_Q^2 + E I_Q + \phi_Q,
\label{eq:mzi_phase}
\end{equation}
Combining these responses, the complex field at the modulator output for mode $m$ is defined as:
\begin{equation}
E_x^{(m)}(I_A, I_Q)
= |E_0^{(m)}|\,
\sqrt{P_{\mathrm{MZI}}(I_A)}\,
\exp\!\big[i\,\phi_{\mathrm{PS}}(I_Q)\big].
\label{eq:iq_field}
\end{equation}

\subsection{Architecture System Design}
\label{Sec_MDT_Arch}

\subsubsection{\textbf{Overall System}}
\label{Sec_MDT_Arch_System}

The architectural system of our proposed \textit{MDTransformer} accelerator is shown in Fig.~\ref{Fig_MDT}(b).
A single \textit{MDTransformer} chip has $N_t$ tiles, and each tile consists of $N_c$ MPTCs. 
An MPTC contains an array of $N_h$$\times$$N_v$ MDOTs.
Furthermore, \textit{MDTransformer} also employs on-chip global SRAM whose size should be at least meeting the minimum required size for storing the largest activations in a layer; following the LT design~\cite{Ref_Zhu_LightTrans_HPCA24}. 
The global SRAM size should not be significantly smaller or larger than this minimum required size, because it can increase the costly off-chip data access (i.e., high access latency and energy) or aggravate the static power consumption, respectively~\cite{Ref_Putra_DRMap_DAC20, Ref_Putra_ROMANet_TVLSI21, Ref_Putra_PENDRAM_arXiv24}.

\smallskip
\subsubsection{\textbf{Dataflow}}
\label{Sec_MDT_Arch_Dataflow}

\begin{figure*}[t]
    \centering
    \includegraphics[width=0.75\linewidth]{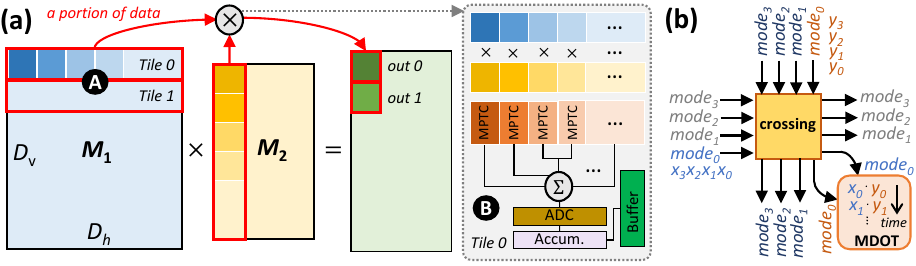}
    \vspace{-0.3cm}
    \caption{Dataflow based on data tiling mechanism for \textit{MDTransformer}. \textbf{(a)} It partitions data from $M_1$ across $D_v$, and map them across tiles. \textbf{(b)} Its crossing routes operands based on their mode to the corresponding MDOT.}
    \label{Fig_Dataflow}
    \vspace{-0.3cm}
\end{figure*}

To maximize benefits of the \textit{MDTransformer} architecture, a specialized dataflow is developed.
Its key ideas are illustrated in Fig.~\ref{Fig_Dataflow} and described below. 
\begin{itemize}[leftmargin=*]
    \item Multiple data is processed in the same MDOT without any prior data programming; see Fig.~\ref{Fig_Dataflow}(a).
    Multiple tiles can process multiple portions of data, which determines the parallelism level in a chip; see~\circledB{A}. 
    Then, multiple cores (MPTCs) can process a portion of data, which also determines the parallelism level in a tile; see~\circledB{B}. 
    Afterward, an $N_h$$\times$$N_v$ MDOT array can perform multiplications in parallel in the core level.     
    \item Each MDOT performs multiplication between two operands from the same mode. 
    Hence, a sequence of multiplications can be scheduled to be performed in the same MDOT, enabling flexible scheduling for exploiting data reuse without expensive broadcast routing; see Fig.~\ref{Fig_Dataflow}(b).
\end{itemize}

\section{Evaluation Methodology}
\label{Sec_EvalMethod}

\begin{figure}[h]
    \vspace{-0.2cm}
    \centering
    \includegraphics[width=\linewidth]{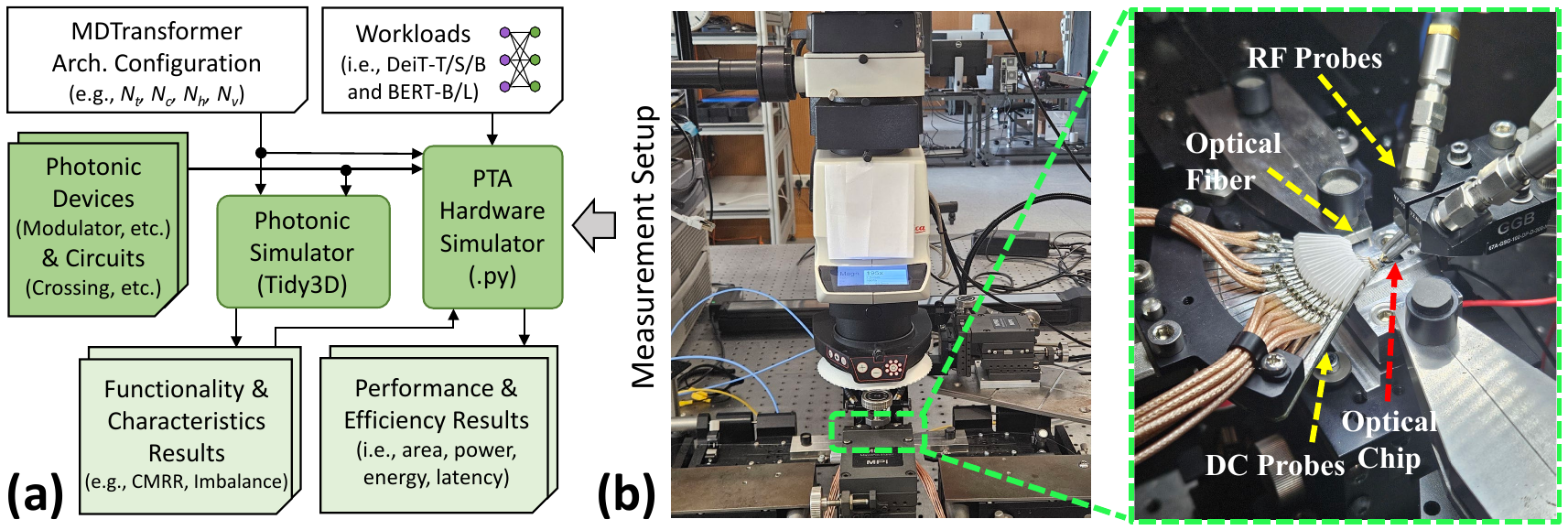}
    \vspace{-0.7cm}
    \caption{\textbf{(a)} Experimental setup and tools flow in this work. \textbf{(b)} Measurement setup for testing the fabricated chip.}
    \label{Fig_TestSetup}
    \vspace{-0.3cm}
\end{figure}

To evaluate our \textit{MDTransformer} design, we employ: (1) functional simulation using Tidy3D~\cite{flexcompute_tidy3d}, and (2) hardware evaluation using the state-of-the-art PTA hardware simulator from~\cite{Ref_Zhu_LightTrans_HPCA24}; see Fig.~\ref{Fig_TestSetup}(a).
We use functional simulation to evaluate the functionality and characteristics of our proposed optical devices and circuits. 
The corresponding results are mainly presented in Section~\ref{Sec_MDT} to validate the functionality of \textit{MDTransformer}.
Meanwhile, we use PTA hardware simulator aims to evaluate area and power of the design as well as its energy consumption and latency when running the workload, while considering device parameters from measurements; see Table~\ref{Table_Params}.
We select DeiT-T, DeiT-S, DeiT-B, BERT-B, and BERT-L as the workloads.
As comparison partners, we use the state-of-the-art LT-Base, LT-Large, and LT-Custom with the following configurations.
\begin{itemize}[leftmargin=*]
    \item \textit{MDTransformer} employs $N_t$=4 tiles, $N_c$=2 cores-per-tile, $N_h$=$N_v$=4 input horizontal/vertical waveguides-per-core, $N_{\lambda}$=1 wavelength, and 4 modes. 
    \item LT-Base employs $N_t$=4, $N_c$=2, and $N_h$=$N_v$=$N_{\lambda}$=12.
    \item LT-Large employs $N_t$=8, $N_c$=2, and $N_h$=$N_v$=$N_{\lambda}$=12.
    \item LT-Custom employs $N_t$=4, $N_c$=2, and $N_h$=$N_v$=$N_{\lambda}$=4.
\end{itemize}
LT-Large employs 4MB global SRAM, while the others use 2MB.
Our design is under fabrication and its measurement setup is shown in Fig.~\ref{Fig_TestSetup}(b).

\begin{table}[h]
\caption{Summary of device parameters}
\label{Table_Params}
\footnotesize
\centering
\renewcommand{\arraystretch}{1.0}
\begin{tabular}{|c|c|c|}
\hline
\textbf{Device} & \textbf{Parameter} & \textbf{Value} \\
\hline
\hline
\multirow{3}{*}{DAC \cite{caragiulo20212}}
 & Precision & 8-bit \\
 & Power & 42 mW (@28 GSPS) \\
 & Area & 0.03 mm$^2$ \\
\hline
\multirow{3}{*}{ADC \cite{duan201412}}
 & Precision & 32 lines, 6-bit \\
 & Power & 410 mW (@12.8 GSPS) \\
 & Area & 780 $\mu$m$^2$ \\
\hline
\multirow{2}{*}{TIA \cite{serunjogi202164gb}}
 & Power & 30 mW \\
 & Area & $<50~\mu$m$^2$ \\
\hline
\multirow{2}{*}{MZM}
 & Power & 50 mW \\
 & Area & $\sim$2{,}260 $\mu$m$^2$ \\
\hline
\multirow{2}{*}{Crossing}
 & IL & 0.3 dB \\
 & Area & 36 $\mu$m$^2$ \\
\hline
\multirow{2}{*}{Phase Shifter}
 & IL & 1 dB \\
 & Area & 250 $\mu$m$^2$ \\
\hline
\multirow{2}{*}{Y-Splitter}
 & IL & 0.4 dB \\
 & Area & 36 $\mu$m$^2$ \\
\hline
\multirow{2}{*}{4 Mode MUX/DEMUX}
 & IL & 6.2 dB \\
 & Area & 64 $\mu$m$^2$ \\
\hline
\multirow{2}{*}{Coherent Hybrid}
 & IL & 6 dB \\
 & Area & 144 $\mu$m$^2$ \\
\hline
\multirow{3}{*}{Photodetector}
 & Power & 1.1 mW \\
 & Sensitivity & -25 dBm \\
 & Area & 4$\times$10$~\mu$m$^2$ \\
\hline
\end{tabular}
\vspace{-0.3cm}
\end{table}

\section{Results and Discussion}
\label{Sec_Results}

\subsection{Reduction of Area and Power Consumption}

\begin{figure*}[t]
    \centering
    \includegraphics[width=\linewidth]{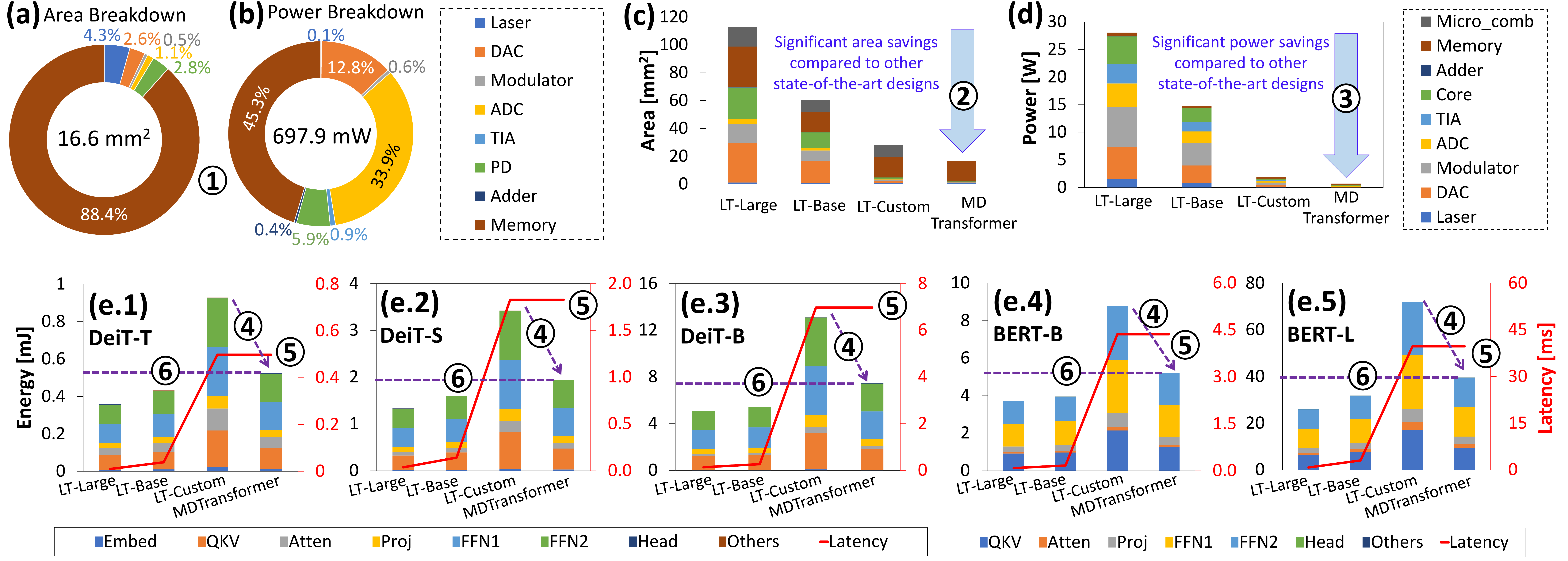}
    \vspace{-0.6cm}
    \caption{Experimental results for \textbf{(a)} area breakdown of MDTransformer; \textbf{(b)} power breakdown of MDTransformer; \textbf{(c)} comparison on area; \textbf{(d)} comparison on power; as well as energy consumption and latency for different workloads: \textbf{(e.1)} DeiT-T, \textbf{(e.2)} DeiT-S, \textbf{(e.3)} DeiT-B; \textbf{(e.4)} BERT-B, and \textbf{(e.5)} BERT-L.}
    \label{Fig_Results}
    \vspace{-0.3cm}
\end{figure*}

Experimental results for area and power consumption are presented in Fig.~\ref{Fig_Results}(a)-(d). 
The results show that \textit{MDTransformer} occupies 16.6mm$^2$ area and incurs 697.9mW power; see \circled{1}. 
These profiles are dominated by on-chip memory as the impact of Micro\_comb, modulator, and phase-shifter is significantly decreased compared to state-of-the-art designs.
The reason is that, our design strategy for developing \textit{MDTransformer} is to eliminate Micro\_comb, reduce modulator size, and remove phase-shifter in MDOT, hence leading to significantly small area and low power consumption. 

\smallskip
\textbf{Area comparison:} 
Our \textit{MDTransformer} significantly saves area compared to all state-of-the-art designs, i.e., reducing area by 85.3\% from LT-Large, 72.4\% from LT-Base, and 40.3\% from LT-Custom; see \circled{2}.
\textit{MDTransformer} occupies smaller area than LT-Custom despite having the same number of tiles, cores, and core size. 
The reason is that, \textit{MDTransformer} employs smaller modulator as well as eliminates Micro\_comb and phase-shifter in MDOT. 
In addition to that, \textit{MDTransformer} also employs smaller number of tiles and smaller core size compared to LT-Large and LT-Base, thus leading to significantly smaller area. 

\smallskip
\textbf{Power comparison:} 
Our \textit{MDTransformer} significantly decreases power compared to all state-of-the-art designs, i.e., reducing power consumption by 97.5\% from LT-Large, 95.3\% from LT-Base, and 63.6\% from LT-Custom; see \circled{3}.
\textit{MDTransformer} incurs smaller power than LT-Custom despite having the same number of tiles, cores, and core size. 
The reason is that, \textit{MDTransformer} employs efficient modulator design as well as completely removes power consumption from Micro\_comb and phase-shifter in MDOT. 
In addition to that, \textit{MDTransformer} also employs smaller number of tiles and smaller core size compared to LT-Large and LT-Base, thus leading to significantly lower power consumption.

\subsection{Enabling High Performance and Energy Efficiency across Transformer Workloads}

Experimental results for energy consumption and latency of core processing on different workloads are provided in Fig.~\ref{Fig_Results}(e.1)-(e.5). 
Based on these results, we make the following key observations.
\begin{itemize}[leftmargin=*]
    \item \textit{MDTransformer} consistently achieves lower energy consumption than LT-Custom across different workloads, despite having the same number of tiles, cores, and core size; as indicated by \circled{4}.
    Specifically, \textit{MDTransformer} saves energy consumption by 43.1\%-43.5\% for DeiT-based models and by 40.6\%-45.1\% for BERT-based models as compared to LT-Custom. 
    The reason is that, \textit{MDTransformer} eliminates power requirement for phase-shifter in MDOT and reduces power cost for modulator, which in turns leading to lower energy consumption when processing the workload.  
    \smallskip
    \item \textit{MDTransformer} achieves comparable processing latency to LT-Custom across different workloads, as shown by \circled{5}.
    The reason is that, these two designs consider the same on-chip memory size as well as the same number of tiles, cores, and core size.
    Therefore, they have similar capabilities in storing data on-chip and performing computation based on their dataflow and scheduling. 
    However, such a similar performance comes at the different cost of power consumption, as shown by \circled{3}.
    Therefore, their energy consumption profiles also differ significantly across different workloads, as indicated by \circled{4}. 
    LT-Large and LT-Base consume relatively low energy since they employ high parallelism to expedite the processing, hence leading to low latency.
    However, this comes at the cost of huge power consumption, as indicated in Fig.~\ref{Fig_Results}(d). 
    This condition may limit the applicability of the LT-Large and LT-Base accelerators for diverse low-power application use-cases.  
    In contrast, \textit{MDTransformer} achieves competitive energy consumption compared to LT-Large and LT-Base, as shown by \circled{6}, while incurring a significantly lower power consumption than LT-Large and LT-Base, as shown by \circled{3}. 
    The reason is that, \textit{MDTransformer} combines the benefits of low-power design through optimized optical devices/modules, selection of architecture configuration, and efficient dataflow for enabling high-performance and energy-efficient optical-based processing.    
\end{itemize}

\vspace{-0.2cm}
\subsection{Further Discussion}

In this work, we consider a configuration of $N_t$=4, $N_c$=2, $N_h$=$N_v$=4, $N_{\lambda}$=1, and 4 modes for our \textit{MDTransformer}. 
However, this selection of configuration can be adjusted based on the requirements.
For instance, if we need to increase the parallelism in the \textit{MDTransformer}, then we can increase the number of tiles $N_t$, number of cores-per-tile $N_c$, and core size $N_h$x$N_v$. 
Conversely, if we have a targeted application that imposes tight design constraints, e.g., in terms of area, power, energy, and latency (or throughput), then the configuration should be selected carefully. 
All these adjustment choices are supported with our dynamically-operated architecture and dataflow design in \textit{MDTransformer}, thereby providing a practical, high-performance and energy-efficient PTA design. 

\section{Conclusion}
\label{Sec_Conclude}

We propose a novel hardware-software co-design of \textit{MDTransformer} accelerator, which employs MDM-based computation, inverse-designed coherent PTC, and IQ modulation.
Experimental results show that, our \textit{MDTransformer} accelerator offers 4-bit effective precision for multiplication, low inter-modal crosstalk (i.e., less than -30dB), and full compatibility with single-laser continuous-wave operation at 1550nm. 
It also saves 40.4\% area, 63.6\% power, and 40.6\% energy consumption, with comparable latency over the state-of-the-art across different transformer models. 
Therefore, our \textit{MDTransformer} accelerator successfully provides a practical solution for high-performance and energy-efficient transformer-based systems.

% ...

\bibliographystyle{IEEEtran}
\bibliography{bibliography}

@article{luo2014wdm,
  title={WDM-compatible mode-division multiplexing on a silicon chip},
  author={Luo, Lian-Wee and Ophir, Noam and Chen, Christine P and Gabrielli, Lucas H and Poitras, Carl B and Bergmen, Keren and Lipson, Michal},
  journal={Nature communications},
  volume={5},
  number={1},
  pages={3069},
  year={2014},
  publisher={Nature Publishing Group UK London}
}

@article{liu2019arbitrarily,
  title={Arbitrarily routed mode-division multiplexed photonic circuits for dense integration},
  author={Liu, Yingjie and Xu, Ke and Wang, Shuai and Shen, Weihong and Xie, Hucheng and Wang, Yujie and Xiao, Shumin and Yao, Yong and Du, Jiangbing and He, Zuyuan and others},
  journal={Nature communications},
  volume={10},
  number={1},
  pages={3263},
  year={2019},
  publisher={Nature Publishing Group UK London}
}

@article{yang2022multi,
  title={Multi-dimensional data transmission using inverse-designed silicon photonics and microcombs},
  author={Yang, Ki Youl and Shirpurkar, Chinmay and White, Alexander D and Zang, Jizhao and Chang, Lin and Ashtiani, Farshid and Guidry, Melissa A and Lukin, Daniil M and Pericherla, Srinivas V and Yang, Joshua and others},
  journal={Nature communications},
  volume={13},
  number={1},
  pages={7862},
  year={2022},
  publisher={Nature Publishing Group UK London}
}

@article{pita2025integrated,
  title={Integrated silicon nitride devices via inverse design},
  author={Pita Ruiz, Julian L and Dalvand, Narges and M{\'e}nard, Micha{\"e}l},
  journal={Nature Communications},
  volume={16},
  number={1},
  pages={9307},
  year={2025},
  publisher={Nature Publishing Group UK London}
}

@article{li2025ultra,
  title={Ultra-compact scalable mode demultiplexers for high-speed optical interconnects via GPU-accelerated inverse design},
  author={Li, Jiahao and Li, Xiang and Wu, Lin and Luo, Ming and Li, Yuan and Wang, Yilun and Qiu, Ying},
  journal={Optics Express},
  volume={33},
  number={21},
  pages={44908--44924},
  year={2025},
  publisher={Optica Publishing Group}
}

@inproceedings{serunjogi202164gb,
  title={64Gb/s NRZ/PAM4 burst-mode optical receiver frontend with gain control, offset correction and gain decoupled from bandwidth},
  author={Serunjogi, Solomon and Rasras, Mahmoud and Sanduleanu, Mihai},
  booktitle={2021 IEEE International Symposium on Circuits and Systems (ISCAS)},
  pages={1--4},
  year={2021},
  organization={IEEE}
}

@article{duan201412,
  title={A 12.8 GS/s time-interleaved ADC with 25 GHz effective resolution bandwidth and 4.6 ENOB},
  author={Duan, Yida and Alon, Elad},
  journal={IEEE Journal of Solid-State Circuits},
  volume={49},
  number={8},
  pages={1725--1738},
  year={2014},
  publisher={IEEE}
}

@article{caragiulo20212,
  title={A 2x Time-Interleaved 28-GS/s 8-Bit 0.03-mm 2 Switched-Capacitor DAC in 16-nm FinFET CMOS},
  author={Caragiulo, Pietro and Mattia, Oscar Elisio and Arbabian, Amin and Murmann, Boris},
  journal={IEEE Journal of Solid-State Circuits},
  volume={56},
  number={8},
  pages={2335--2346},
  year={2021},
  publisher={IEEE}
}

@article{lam2024dynamic,
  title={Dynamic electro-optic analog memory for neuromorphic photonic computing},
  author={Lam, Sean and Khaled, Ahmed and Bilodeau, Simon and Marquez, Bicky A and Prucnal, Paul R and Chrostowski, Lukas and Shastri, Bhavin J and Shekhar, Sudip},
  journal={arXiv preprint arXiv:2401.16515},
  year={2024}
}

@article{ning2024photonic,
  title={Photonic-electronic integrated circuits for high-performance computing and ai accelerators},
  author={Ning, Shupeng and Zhu, Hanqing and Feng, Chenghao and Gu, Jiaqi and Jiang, Zhixing and Ying, Zhoufeng and Midkiff, Jason and Jain, Sourabh and Hlaing, May H and Pan, David Z and others},
  journal={Journal of Lightwave Technology},
  year={2024},
  publisher={IEEE}
}

@article{babashah2019temporal,
  title={Temporal analog optical computing using an on-chip fully reconfigurable photonic signal processor},
  author={Babashah, Hossein and Kavehvash, Zahra and Khavasi, Amin and Koohi, Somayyeh},
  journal={Optics \& Laser Technology},
  volume={111},
  pages={66--74},
  year={2019},
  publisher={Elsevier}
}

@article{rahimi2024realization,
  title={Realization of an integrated coherent photonic platform for scalable matrix operations},
  author={Rahimi Kari, Sadra and Nobile, Nicholas A and Pantin, Dominique and Shah, Vivswan and Youngblood, Nathan},
  journal={Optica},
  volume={11},
  number={4},
  pages={542--551},
  year={2024},
  publisher={Optica Publishing Group}
}

@article{wang2022mode,
  title={Mode division multiplexing on an InP membrane on silicon},
  author={Wang, Yi and Wei, Yihui and Dolores-Calzadilla, Victor and Williams, Kevin and Smit, Meint and Dai, Daoxin and Jiao, Yuqing},
  journal={Optics Letters},
  volume={47},
  number={16},
  pages={4004--4007},
  year={2022},
  publisher={Optica Publishing Group}
}

@misc{flexcompute_tidy3d,
  author       = {Flexcompute, Inc.},
  title        = {{Tidy3D}: Next-generation electromagnetic simulation tool},
  howpublished = {\url{https://www.flexcompute.com/tidy3d/solver/}},
  year         = {2024},
  note         = {Accessed: 2025-01-01}
}

@article{banerjee2022characterizing,
  title={Characterizing coherent integrated photonic neural networks under imperfections},
  author={Banerjee, Sanmitra and Nikdast, Mahdi and Chakrabarty, Krishnendu},
  journal={Journal of lightwave technology},
  volume={41},
  number={5},
  pages={1464--1479},
  year={2022},
  publisher={IEEE}
}

@article{totovic2022programmable,
  title={Programmable photonic neural networks combining WDM with coherent linear optics},
  author={Totovic, Angelina and Giamougiannis, George and Tsakyridis, Apostolos and Lazovsky, David and Pleros, Nikos},
  journal={Scientific reports},
  volume={12},
  number={1},
  pages={5605},
  year={2022},
  publisher={Nature Publishing Group UK London}
}

@article{biasi2024photonic,
  title={Photonic neural networks based on integrated silicon microresonators},
  author={Biasi, Stefano and Donati, Giovanni and Lugnan, Alessio and Mancinelli, Mattia and Staffoli, Emiliano and Pavesi, Lorenzo},
  journal={Intelligent Computing},
  volume={3},
  pages={0067},
  year={2024},
  publisher={AAAS}
}

@article{tait2016microring,
  title={Microring weight banks},
  author={Tait, Alexander N and Wu, Allie X and De Lima, Thomas Ferreira and Zhou, Ellen and Shastri, Bhavin J and Nahmias, Mitchell A and Prucnal, Paul R},
  journal={IEEE Journal of Selected Topics in Quantum Electronics},
  volume={22},
  number={6},
  pages={312--325},
  year={2016},
  publisher={IEEE}
}

@article{hamerly2024netcast,
  title={Netcast: low-power edge computing with WDM-defined optical neural networks},
  author={Hamerly, Ryan and Sludds, Alexander and Bandyopadhyay, Saumil and Chen, Zaijun and Zhong, Zhizhen and Bernstein, Liane and Englund, Dirk},
  journal={Journal of Lightwave Technology},
  volume={42},
  number={22},
  pages={7795--7806},
  year={2024},
  publisher={IEEE}
}

@article{li2025hybrid,
  title={Hybrid Photonic-digital Accelerator for Attention Mechanism},
  author={Li, Huize and Chen, Dan and Mitra, Tulika},
  journal={arXiv preprint arXiv:2501.11286},
  year={2025}
}

@article{sapra2020inverse,
  title={Inverse design of compact multimode multi-port photonic devices},
  author={Sapra, Neil V. et al.},
  journal={Nature Communications},
  volume={11},
  pages={6361},
  year={2020}
}

@article{jensen2020inverse,
  title={Adjoint-based inverse design of efficient, broadband mode conversion devices},
  author={Jensen, Jacob S. et al.},
  journal={ACS Photonics},
  volume={7},
  pages={1497--1506},
  year={2020}
}

@article{vercruysse2020compact,
  title={Compact broadband directional couplers using inverse design},
  author={Vercruysse, Dries et al.},
  journal={Optica},
  volume={7},
  pages={179--185},
  year={2020}
}

@String{BIT = "{BIT}" }

@String{Computing = "Computing" }

@String{Computer = "{IEEE} Computer" }

@ARTICLE{Ref_Yenduri_AGIsurvey_Access25,
  author={Yenduri, Gokul and Murugan, Ramalingam and Kumar Reddy Maddikunta, Praveen and Bhattacharya, Sweta and Sudheer, Devulapalli and Bhushan Savarala, Bharath},
  journal={IEEE Access}, 
  title={Artificial General Intelligence: Advancements, Challenges, and Future Directions in AGI Research}, 
  year={2025},
  volume={13},
  number={},
  pages={134325-134356},
  }

@INPROCEEDINGS{Ref_Li_SPACX_HPCA22,
  author={Li, Yuan and Louri, Ahmed and Karanth, Avinash},
  booktitle={2022 IEEE International Symposium on High-Performance Computer Architecture (HPCA)}, 
  title={SPACX: Silicon Photonics-based Scalable Chiplet Accelerator for DNN Inference}, 
  year={2022},
  volume={},
  number={},
  pages={831-845},
  doi={10.1109/HPCA53966.2022.00066}}

@ARTICLE{Ref_Li_SPRINT_TPDS22,
  author={Li, Yuan and Louri, Ahmed and Karanth, Avinash},
  journal={IEEE Transactions on Parallel and Distributed Systems (TPDS)}, 
  title={SPRINT: A High-Performance, Energy-Efficient, and Scalable Chiplet-Based Accelerator With Photonic Interconnects for CNN Inference}, 
  year={2022},
  volume={33},
  number={10},
  pages={2332-2345},
  }

@inproceedings{Ref_Yin_Simphony_DAC25,
  title={Simphony: A device-circuit-architecture cross-layer modeling and simulation framework for heterogeneous electronic-photonic ai system},
  author={Yin, Ziang and Zhang, Meng and Gangi, Nicholas and Huang, Rena and Zhang, Jeff and Gu, Jiaqi},
  booktitle={2025 62nd ACM/IEEE Design Automation Conference (DAC)},
  pages={1--7},
  year={2025},
  organization={IEEE}
}

@ARTICLE{Ref_Li_MERIT_TSUSC25,
  author={Li, Yuan and Louri, Ahmed and Karanth, Avinash},
  journal={IEEE Transactions on Sustainable Computing (TSUSC)}, 
  title={MERIT: A Sustainable DNN Accelerator Design With Photonic Phase-Change Memory}, 
  year={2025},
  volume={10},
  number={4},
  pages={705-716},
  }

@inproceedings{Ref_Afifi_LLMoptical_GLSVLSI25,
  title={A Light-Speed Large Language Model Accelerator with Optical Stochastic Computing},
  author={Afifi, Salma and Alo, Oluwaseun and Thakkar, Ishan and Pasricha, Sudeep},
  booktitle={Great Lakes Symposium on VLSI (GLSVLSI) 2025},
  year={2025}
}

@INPROCEEDINGS{Ref_Li_HyAtten_DATE25,
  author={Li, Huize and Chen, Dan and Mitra, Tulika},
  booktitle={2025 Design, Automation \& Test in Europe Conference (DATE)}, 
  title={HyAtten: Hybrid Photonic-Digital Architecture for Accelerating Attention Mechanism}, 
  year={2025},
  volume={},
  number={},
  pages={1-7},
  doi={10.23919/DATE64628.2025.10993031}}

@article{Ref_Afifi_ASTRA_TECS25,
  title={ASTRA: A Stochastic Transformer Neural Network Accelerator with Silicon Photonics},
  author={Afifi, Salma and Alo, Oluwaseun and Thakkar, Ishan and Pasricha, Sudeep},
  journal={ACM Transactions on Embedded Computing Systems (TECS)},
  year={2025},
  publisher={ACM New York, NY}
}

@article{Ref_Sunny_SurveyPhot4DL_JETC21,
  title={A survey on silicon photonics for deep learning},
  author={Sunny, Febin P and Taheri, Ebadollah and Nikdast, Mahdi and Pasricha, Sudeep},
  journal={ACM Journal of Emerging Technologies in Computing System (JETC)},
  volume={17},
  number={4},
  pages={1--57},
  year={2021},
  publisher={ACM New York, NY}
}

@inproceedings{Ref_Afifi_TRON_GLSVLSI23,
  title={Tron: Transformer neural network acceleration with non-coherent silicon photonics},
  author={Afifi, Salma and Sunny, Febin and Nikdast, Mahdi and Pasricha, Sudeep},
  booktitle={Great Lakes Symposium on VLSI (GSVLSI) 2023},
  pages={15--21},
  year={2023}
}

@article{Ref_Zhu_ENlighten_arXiv25,
  title={ENLighten: Lighten the Transformer, Enable Efficient Optical Acceleration},
  author={Zhu, Hanqing and Zhou, Zhican and Ning, Shupeng and Wu, Xuhao and Chen, Ray and Wan, Yating and Pan, David},
  journal={arXiv preprint arXiv:2510.01673},
  year={2025}
}

@INPROCEEDINGS{Ref_Chang_PDAC_DAC25,
  author={Chang, Wen-Tse and Wu, Chun-Feng and Lo, Yun-Chen},
  booktitle={2025 62nd ACM/IEEE Design Automation Conference (DAC)}, 
  title={P-DAC: Power-Efficient Photonic Accelerators for LLM Inference}, 
  year={2025},
  volume={},
  number={},
  pages={1-7},
  doi={10.1109/DAC63849.2025.11132618}}

@inproceedings{Ref_Shiflett_Albireo_ISCA21,
  title={Albireo: Energy-efficient acceleration of convolutional neural networks via silicon photonics},
  author={Shiflett, Kyle and Karanth, Avinash and Bunescu, Razvan and Louri, Ahmed},
  booktitle={2021 ACM/IEEE 48th Annual International Symposium on Computer Architecture (ISCA)},
  pages={860--873},
  year={2021},
  organization={IEEE}
}

@article{Ref_Feldmann_PCMcrossbar_Nature21,
  title={Parallel convolutional processing using an integrated photonic tensor core},
  author={Feldmann, Johannes and Youngblood, Nathan and Karpov, Maxim and Gehring, Helge and Li, Xuan and Stappers, Maik and Le Gallo, Manuel and Fu, Xin and Lukashchuk, Anton and Raja, Arslan S and others},
  journal={Nature},
  volume={589},
  number={7840},
  pages={52--58},
  year={2021},
  publisher={Nature Publishing Group UK London}
}

@article{Ref_Tait_PhotWeightBanks_SciRep17,
  title={Neuromorphic photonic networks using silicon photonic weight banks},
  author={Tait, Alexander N and De Lima, Thomas Ferreira and Zhou, Ellen and Wu, Allie X and Nahmias, Mitchell A and Shastri, Bhavin J and Prucnal, Paul R},
  journal={Scientific Reports},
  volume={7},
  number={1},
  pages={7430},
  year={2017},
  publisher={Nature Publishing Group UK London}
}

@inproceedings{Ref_Sunny_CrossLight_DAC21,
  title={CrossLight: A cross-layer optimized silicon photonic neural network accelerator},
  author={Sunny, Febin and Mirza, Asif and Nikdast, Mahdi and Pasricha, Sudeep},
  booktitle={2021 58th ACM/IEEE design automation conference (DAC)},
  pages={1069--1074},
  year={2021},
  organization={IEEE}
}

@article{Ref_Shen_DLwCoherentPhot_NaturePhot17,
  title={Deep learning with coherent nanophotonic circuits},
  author={Shen, Yichen and Harris, Nicholas C and Skirlo, Scott and Prabhu, Mihika and Baehr-Jones, Tom and Hochberg, Michael and Sun, Xin and Zhao, Shijie and Larochelle, Hugo and Englund, Dirk and others},
  journal={Nature photonics},
  volume={11},
  number={7},
  pages={441--446},
  year={2017},
  publisher={Nature Publishing Group UK London}
}

@article{Ref_Shastri_Photonics4AI_NaturePhot21,
  title={Photonics for artificial intelligence and neuromorphic computing},
  author={Shastri, Bhavin J and others},
  journal={Nature Photonics},
  volume={15},
  number={2},

  year={2021},
  publisher={Nature Publishing Group UK London}
}

@inproceedings{Ref_Zhou_Transpim_HPCA22,
  title={Transpim: A memory-based acceleration via software-hardware co-design for transformer},
  author={Zhou, Minxuan and Xu, Weihong and Kang, Jaeyoung and Rosing, Tajana},
  booktitle={2022 IEEE International Symposium on High-Performance Computer Architecture (HPCA)},
  pages={1071--1085},
  year={2022},
  organization={IEEE}
}

@inproceedings{Ref_You_Vitcod_HPCA23,
  title={Vitcod: Vision transformer acceleration via dedicated algorithm and accelerator co-design},
  author={You, Haoran and Sun, Zhanyi and Shi, Huihong and Yu, Zhongzhi and Zhao, Yang and Zhang, Yongan and Li, Chaojian and Li, Baopu and Lin, Yingyan},
  booktitle={2023 IEEE International Symposium on High-Performance Computer Architecture (HPCA)},
  pages={273--286},
  year={2023},
  organization={IEEE}
}

@inproceedings{Ref_Wang_Spatten_HPCA21,
  title={Spatten: Efficient sparse attention architecture with cascade token and head pruning},
  author={Wang, Hanrui and Zhang, Zhekai and Han, Song},
  booktitle={2021 IEEE International Symposium on High-Performance Computer Architecture (HPCA)},
  pages={97--110},
  year={2021},
  organization={IEEE}
}

@INPROCEEDINGS{Ref_Zhu_LightTrans_HPCA24,
  author={Zhu, Hanqing and Gu, Jiaqi and Wang, Hanrui and Jiang, Zixuan and Zhang, Zhekai and Tang, Rongxing and Feng, Chenghao and Han, Song and Chen, Ray T. and Pan, David Z.},
  booktitle={2024 IEEE International Symposium on High-Performance Computer Architecture (HPCA)}, 
  title={Lightening-Transformer: A Dynamically-Operated Optically-Interconnected Photonic Transformer Accelerator}, 
  year={2024},
  volume={},
  number={},
  pages={686-703},
  doi={10.1109/HPCA57654.2024.00059}}

@inproceedings{Ref_Touvron_Transformers_ICML21,
  title={Training data-efficient image transformers \& distillation through attention},
  author={Touvron, Hugo and Cord, Matthieu and Douze, Matthijs and Massa, Francisco and Sablayrolles, Alexandre and J{\'e}gou, Herv{\'e}},
  booktitle={International Conference on Machine Learning (ICML)},
  pages={10347--10357},
  year={2021},
  organization={PMLR}
}

@article{Ref_Khan_SurveyViT_CSUR22,
  title={Transformers in Vision: A Survey},
  author={Khan, Salman and Naseer, Muzammal and Hayat, Munawar and Zamir, Syed Waqas and Khan, Fahad Shahbaz and Shah, Mubarak},
  journal={ACM Computing Surveys (CSUR)},
  volume={54},
  number={10s},
  pages={1--41},
  year={2022},
  publisher={ACM New York, NY}
}

@article{Ref_Han_SurveyViT_TPAMI22,
  author={Han, Kai and Wang, Yunhe and Chen, Hanting and Chen, Xinghao and Guo, Jianyuan and Liu, Zhenhua and Tang, Yehui and Xiao, An and Xu, Chunjing and Xu, Yixing and Yang, Zhaohui and Zhang, Yiman and Tao, Dacheng},
  journal={IEEE Transactions on Pattern Analysis and Machine Intelligence (TPAMI)}, 
  title={A Survey on Vision Transformer}, 
  year={2023},
  volume={45},
  number={1},
  pages={87-110},
  doi={10.1109/TPAMI.2022.3152247}}

@inproceedings{Ref_Dosovitskiy_Transformers_ICLR21,
title={An Image is Worth 16x16 Words: Transformers for Image Recognition at Scale},
author={Alexey Dosovitskiy and Lucas Beyer and Alexander Kolesnikov and Dirk Weissenborn and Xiaohua Zhai and Thomas Unterthiner and Mostafa Dehghani and Matthias Minderer and Georg Heigold and Sylvain Gelly and Jakob Uszkoreit and Neil Houlsby},
booktitle={International Conference on Learning Representations (ICLR)},
year={2021},
}

@article{Ref_Vaswani_Attention_NIPS17,
  title={Attention is all you need},
  author={Vaswani, Ashish and Shazeer, Noam and Parmar, Niki and Uszkoreit, Jakob and Jones, Llion and Gomez, Aidan N and others},
  journal={Advances in Neural Information Processing Systems (NIPS)},
  volume={30},
  number={1},
  pages={261--272},
  year={2017},}

@ARTICLE{Ref_Putra_ROMANet_TVLSI21,
  author={Putra, Rachmad Vidya Wicaksana and Hanif, Muhammad Abdullah and Shafique, Muhammad},
  journal={IEEE Transactions on Very Large Scale Integration Systems (TVLSI)},
  title={ROMANet: Fine-Grained Reuse-Driven Off-Chip Memory Access Management and Data Organization for Deep Neural Network Accelerators}, 
  year={2021},
  volume={29},
  number={4},
  pages={702-715},
  doi={10.1109/TVLSI.2021.3060509}}

@article{Ref_Putra_PENDRAM_arXiv24,
  title={PENDRAM: Enabling High-Performance and Energy-Efficient Processing of Deep Neural Networks through a Generalized DRAM Data Mapping Policy},
  author={Putra, Rachmad Vidya Wicaksana and Hanif, Muhammad Abdullah and Shafique, Muhammad},
  journal={arXiv preprint arXiv:2408.02412},
  year={2024}
}

@INPROCEEDINGS{Ref_Putra_DRMap_DAC20,
  author={Putra, Rachmad Vidya Wicaksana and Hanif, Muhammad Abdullah and Shafique, Muhammad},
  booktitle={2020 57th ACM/IEEE Design Automation Conference (DAC)}, 
  title={DRMap: A Generic DRAM Data Mapping Policy for Energy-Efficient Processing of Convolutional Neural Networks}, 
  year={2020},
  volume={},
  number={},
  pages={1-6},
  doi={10.1109/DAC18072.2020.9218672}}

@ArtifactSoftware{R,
    title = {R: A Language and Environment for Statistical Computing},
    author = {{R Core Team}},
    organization = {R Foundation for Statistical Computing},
    address = {Vienna, Austria},
    year = {2019},
    url = {https://www.R-project.org/},
}

\end{document}